\documentclass[11pt]{article}
\usepackage{amsmath,amssymb}
\usepackage{multicol,citesort,epsfig}
\begin{document}

\title{Phase separation in mixtures of
colloids and long ideal polymer coils}

\author{{\bf Richard P. Sear}\\
~\\
Department of Physics, University of Surrey,\\
Guildford, Surrey GU2 7XH, United Kingdom\\
email: r.sear@surrey.ac.uk}
%tel. +44 (0)1483 876793\\
%fax +44 (0)1483 876781}
%\date{}

\maketitle

\begin{abstract}
Colloidal suspensions with free polymer coils which are larger
than the colloidal particles are considered. The polymer-colloid
interaction is modeled by an extension of the Asakura-Oosawa model.
Phase separation occurs into dilute and dense fluid phases of colloidal
particles when polymer is added. The critical density of this
transition tends to zero as the size of the polymer coils diverges.
\end{abstract}

%\newpage
\begin{multicols}{2}

By the addition of nonadsorbing polymer,
colloidal suspensions can be made to separate into a dilute phase,
a colloidal `vapour', and a dense phase, a colloidal `liquid'
\cite{poon95}. Nonadsorbing polymer does not adsorb
onto the surfaces of the particles, the interaction between a
monomer and colloidal particle is repulsive.
The colloidal particles are spheres of diameter $\sigma$, and
we can characterise the size of a polymer coil with its
root-mean-square end-to-end separation, $R_E$.
For polymer coils smaller than the particles, $R_E<\sigma$,
the polymer-colloid interaction is,
to a not unreasonable first approximation,
a hard-sphere repulsion: the Asakura-Oosawa model \cite{asakura54,vrij76}.
For values of $R_E/\sigma=O(0.1)$, it has been found that to a good
approximation the effect of the polymer is to create a short-ranged,
range ${\cal R}\simeq\sigma+R_E$, pairwise additive attraction
between the colloidal particles
\cite{meijer94,dijkstra99}. If this range is not too short
it induces a phase separation into dilute and dense fluid phases
of the colloidal particles.
In the other limit, that of
large values of the ratio $R_E/\sigma$, it is known that the
polymer creates an effective attraction between the colloidal
particles but that this is not pairwise additive \cite{meijer94,hanke99}.
Here, we extend the Asakura-Oosawa model to deal with
values of $R_E/\sigma>1$, and then go on to
show that the phase behaviour for large $R_E/\sigma$ is
very different from that induced by a pairwise additive attraction.
As $R_E\rightarrow\infty$ the colloid density
at the critical point tends to zero, whereas with
a pairwise additive attraction, however long the range
of the attraction ${\cal R}$, the density
at the critical point remains nonzero.
A van der
Waals fluid has a pairwise additive attraction of infinite range
but a critical volume fraction of $0.13$ \cite{vdw}.
The colloidal particles can be smaller than the polymer coils
either because the particles are small,
only a few nanometers across, the typical size of
a globular protein \cite{kulkarni99},
or the polymer coils are large, as they are for DNA \cite{verma98}.

\begin{figure}
\begin{center}
\caption{
\lineskip 2pt
\lineskiplimit 2pt
A schematic of the extended Asakura-Oosawa model. The black discs
represent the colloids and the curve represents a polymer coil.
The rescaled monomers used to estimate the polymer-colloid interaction
are drawn as dashed circles.
\label{figschem}
}
\vspace*{0.1in}
\epsfig{file=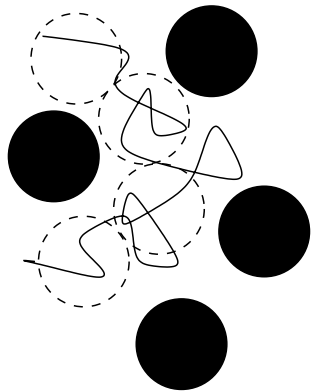,width=1.6in}
\end{center}
\end{figure}

The colloidal particles are modeled by hard spheres, there
are no attractions between them. The interaction potential between
two colloids, $u_{CC}$, is then
\begin{equation}
u_{CC}(r)=\left\{
\begin{array}{ll}
\infty & ~~~~~~ r<\sigma\\
0 & ~~~~~~ r\ge 0
\end{array}\right. ,
\label{ucc}
\end{equation}
where $r$ is the distance between their centres. The polymer coils
are taken to be ideal and so do not interact with each other, i.e.,
$u_{PP}(r)=0$. For the interaction between a colloidal particle and
a polymer coil, $u_{CP}$, we start from the
Asakura-Oosawa model \cite{asakura54,vrij76}
\begin{equation}
u_{CP}(r)=\left\{
\begin{array}{ll}
\infty & ~~~~~~ r<(\sigma+\sigma_P)/2\\
0 & ~~~~~~ r \ge(\sigma+\sigma_P)/2
\end{array}\right. ,
\label{ucp}
\end{equation}
where $\sigma_P$ is an effective diameter of the polymer coil; 
it is close to $R_E$, the root-mean-square end-to-end separation
of the polymer.
The polymer does not adsorb onto the particles; the interaction
between a monomer and the surface of a particle is repulsive.
When the polymer is no larger than the colloid the
Asakura-Oosawa model is reasonable. In the limit $\sigma\gg R_E$,
the colloidal particle resembles a hard planar wall on the relevant length
scale for the polymer, $R_E$, and a hard wall excludes a polymer from
a slab of height of order $R_E$ \cite{eisenriegler96,casassa97}.
However, in the opposite limit, that of $R_E\gg\sigma$, the Asakura-Oosawa
model is incorrect, it predicts that a particle excludes a polymer
coil from a volume of $R_E^3$, whereas if $R_E$ is much larger than
the diameter of the colloid the colloid-polymer interaction must
be extensive in the length of the polymer. For an ideal polymer
$R_E=an^{1/2}$, where $a$ is the monomer length and $n$ is the number
of monomers. The interaction must be extensive in $n$ and so scales
as $R_E^2$ not $R_E^3$ \cite{eisenriegler96,casassa97}.
Thus, we cannot use the Asakura-Oosawa
model for long polymers. We propose an extended Asakura--Oosawa model
to deal with the case $R_E>\sigma$. Figure \ref{figschem} is a schematic
of the proposed model.

First we rescale the monomer size
to the colloid diameter $\sigma$ \cite{degennes}.
As both $R_E$ and the exponent of one half remain constant when we
rescale the monomer size, we have $R_E^2=a^2n=\sigma^2n_b$, where
$n_b$ is the number of blobs: effective monomers of length $\sigma$.
This yields
\begin{equation}
n_b=R_E^2/\sigma^2, ~~~~~n_B\ge 1.
\label{nb}
\end{equation}
So, we now have a polymer of $n_b$ blobs, each of which is $\sigma$
across. Each of these blobs is no larger than the colloidal particle
so the Asakura-Oosawa interaction, Eq. (\ref{ucp}), is a reasonable
(although not a quantitative) description of the interaction of
a single blob with a colloid. Thus, our model for a long polymer is
an ideal chain of blobs of diameter $\sigma$, each of which
interacts with a colloidal particle with an interaction potential given
by Eq. (\ref{ucp}) with $\sigma_p=\sigma$.

For ideal polymers, when calculating phase diagrams,
it is simplest to work in a semigrand ensemble
\cite{lekkerkerker92,meijer94,frenkel00}. This is an ensemble in which
the variables are the volume, $V$, the number of colloidal particles, $N_C$,
and the activity of the polymer, $z$. As all our interactions
are athermal, the temperature $T$ is a not a relevant variable.
For simplicity we will use units such that the thermal energy $kT=1$.
The number density of the colloidal particles $\rho_C=N_C/V$.
We use a reduced density for the colloidal particles, the volume
fraction $\eta=(\pi/6)(N_C/V)\sigma^3$, and a reduced activity
of the polymer $z^*=zR_E^3$.
We also define a reduced density of polymer, $\rho^*=\rho_PR_E^3$,
where $\rho_P$ is the number density of polymer coils.
The colloidal volume fraction
is close to one half when the colloid (in the absence of polymer)
crystallises, and $\rho^*$ is close to one when the polymer coils start to
overlap, and is less than one in the dilute regime \cite{degennes}.

The semigrand potential per colloidal particle $\omega$ is
\cite{lekkerkerker92,meijer94,poon95,frenkel00,dijkstra99}
\begin{equation}
\omega(\eta,z)= a_{HS}(\eta)-(z/\rho_C)\alpha(\eta;n_B),
\label{adef}
\end{equation}
where $a_{HS}$ is the Helmholtz free energy per particle of hard spheres,
and $\alpha(\eta;n_B)=\exp\left[-\mu_{EX}(\eta;n_B)\right]$,
with $\mu_{EX}(\eta,n_B)$ the excess chemical potential of a chain
of $n_B$ blobs in a system of hard spheres at a volume fraction $\eta$.
Equation (\ref{adef}) for the semigrand potential is approximate,
it is essentially the semigrand potential expressed as the $z=0$
limit (no polymer)
plus a series expansion in powers of $z$, truncated after the
first term. Thus, it becomes less accurate as the polymer activity
increases. It can be derived in a couple of ways, see Refs.
\cite{lekkerkerker92,meijer94,poon95,frenkel00,dijkstra99}.
The Carnahan-Starling equation
of state \cite{carnahan69} is known to be accurate so we will use
thermodynamic functions and correlation functions derived from it.
The density of polymer coils when its activity is $z^*$ is
\begin{equation}
\rho^*=z^*\alpha(\eta,n_B),
\label{rhos}
\end{equation}
which is just the definition of the excess chemical potential,
rearranged.

If $n_B=1$, then the polymer-colloid interaction
is modeled by just one hard sphere of diameter $\sigma$, as in the
original Asakura-Oosawa model. Then $\mu_{EX}$ is just the excess
chemical potential of hard spheres, which may be easily derived
from the Carnahan-Starling equation of state. For $n_B>1$ we
require the excess chemical potential of a chain of blobs. Fortunately
the problem of calculating this quantity has occurred in the treatment
of dense liquids of oligomers, such as alkanes, and polymer melts.
One of the best known theories is that of
Wertheim \cite{wertheim87}, which he termed
thermodynamic perturbation theory 1 (TPT1).
Its prediction for $\mu_{EX}(\eta,n_B)$ is \cite{wertheim87,sear94,wnote}
\begin{eqnarray}
\mu_{EX}= n_B\mu_{EX}^{(HS)}-
\left(n_B-1\right)\left[\ln g_{HS}+\frac{\eta}{g_{HS}}
\left(\frac{{\rm d}g_{HS}}{{\rm d}\eta}\right)\right],\nonumber\\
\label{muex}
\end{eqnarray}
where $\mu_{EX}^{(HS)}$ and $g_{HS}$, are the excess chemical
potential and pair distribution function at contact, respectively,
of hard spheres. $\mu_{EX}$ is the work done in inserting
a chain of $n_B$ spheres of diameter $\sigma$ into the fluid of
hard spheres,
which is equal to the work done in inserting $n_B$ widely separated
spheres (the first term on the right hand side of Eq. (\ref{muex})
plus the work done in bringing the $n_B$ spheres together
into a linear chain of spheres at contact (the second term) \cite{wnote}.

\begin{figure}
\begin{center}
\caption{
\lineskip 2pt
\lineskiplimit 2pt
The phase diagram of a colloid + polymer mixture, with a polymer
of $n_B=5$ blobs of diameter equal to that of the colloid.
a) is the diagram in the $\eta-z^*$ plane, and b)
is it in the $\eta-\rho^*$ plane.
The dashed lines are tie
lines connecting coexisting phases and the circle in b) marks the
critical point.
\label{figrho}
}
\vspace*{0.1in}
\epsfig{file=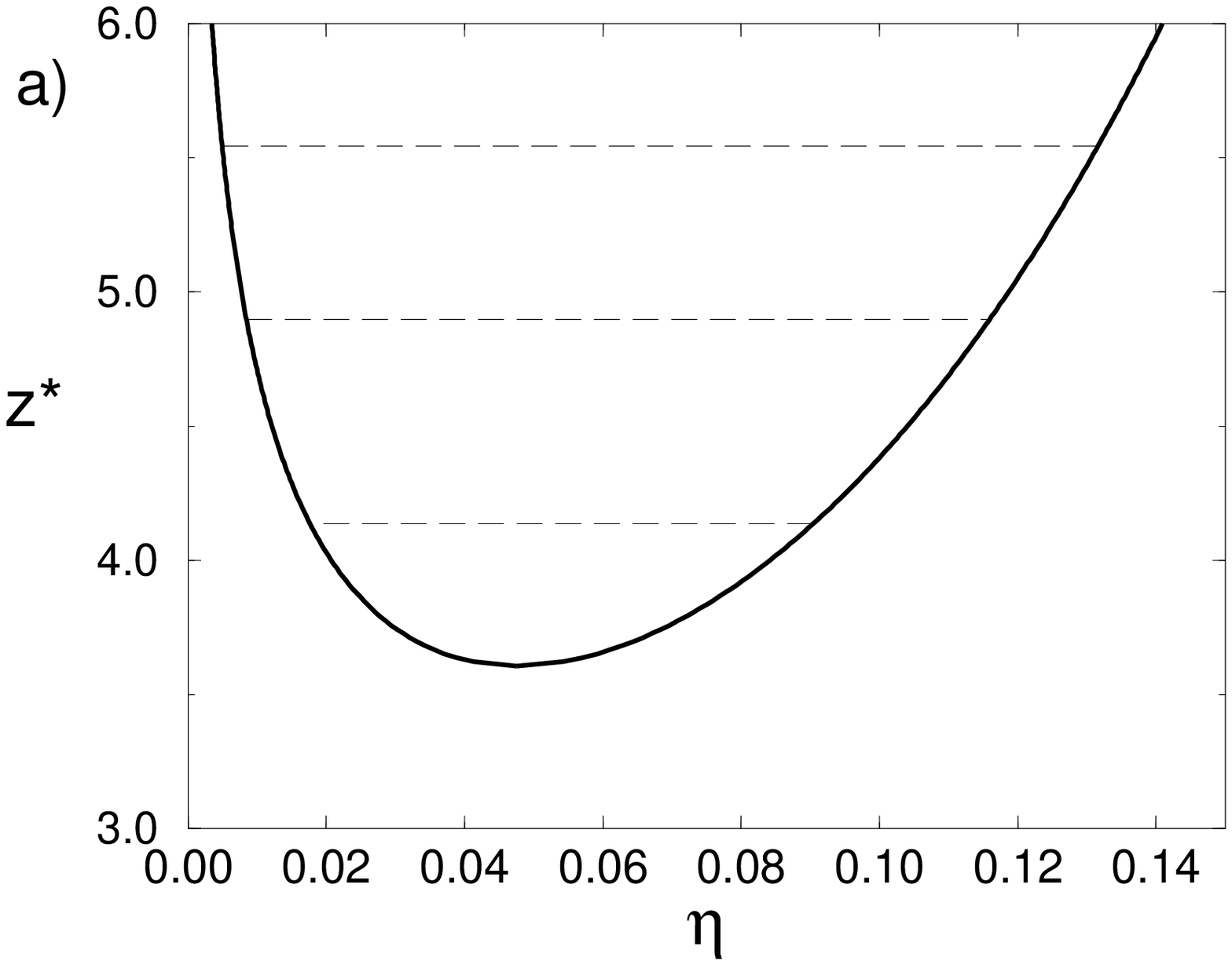,width=3.0in}
%\end{center}
%\begin{center}
%\vspace*{0.1in}
\epsfig{file=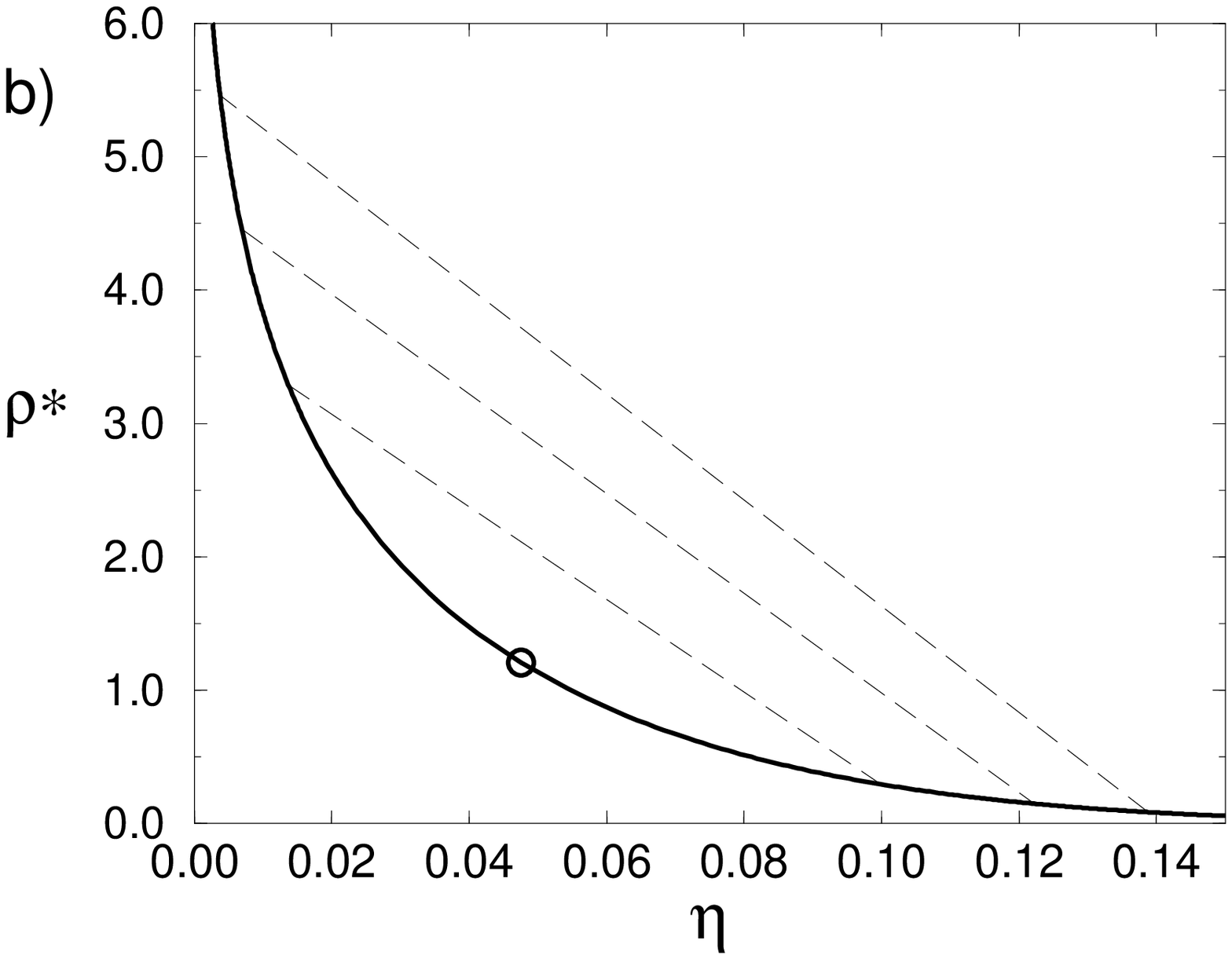,width=3.0in}
\end{center}
\end{figure}

Equations (\ref{adef}) and (\ref{muex}) are all that
is required to calculate the phase diagram in the $\eta-z^*$
plane.
Then  Eq. (\ref{rhos}) can be used to calculate the polymer
density from its activity and so these diagrams may be mapped
onto the $\eta-\rho^*$ plane.
The pressure and the
chemical potential of the colloid are both derivatives of $\Omega$, and so
may be determined from Eq. (\ref{adef}), and then the conditions
of equal chemical potential and pressure may be solved for phase
coexistence. Results for $n_B=5$ are shown in Fig. \ref{figrho}.
This corresponds to $R_E=\sqrt{5}\sigma$, the end-to-end
separation of the polymer is a little over twice the diameter of the
colloid. If the colloid is a globular protein molecule this
corresponds to an $R_E$ of around 10nm.

The density of polymer coils
decreases exponentially with increasing colloid volume fraction,
with a coefficient in the exponential
which is linear in $n_B$, Eq. (\ref{muex}). Thus
for polymers which are several blobs long, the polymer density
at a polymer activity $z^*=O(1)$ and at high colloid volume
fractions, $\eta\gtrsim0.3$, is extremely small. Thus, crystallisation
of the colloid takes place in the presence of almost no polymer and
so occurs very close to its value for hard spheres, which is
$\eta=0.49$ \cite{hoover68},
so is almost completely unaffected at these polymer activities.
We do not show the crystallisation transition in Fig. \ref{figrho}
because it is at much higher densities than the vapour-liquid transition
but the density of the fluid phase which coexists with the crystalline
phase is essentially a vertical line at a volume fraction of $0.49$.

\begin{figure}
\begin{center}
\caption{
\lineskip 2pt
\lineskiplimit 2pt
The volume fraction of colloidal particles at the fluid-fluid
critical point, $\eta_{CP}$, as a function
of the size of the polymer, measured by $n_B$.
\label{figcps}
}
\vspace*{0.1in}
\epsfig{file=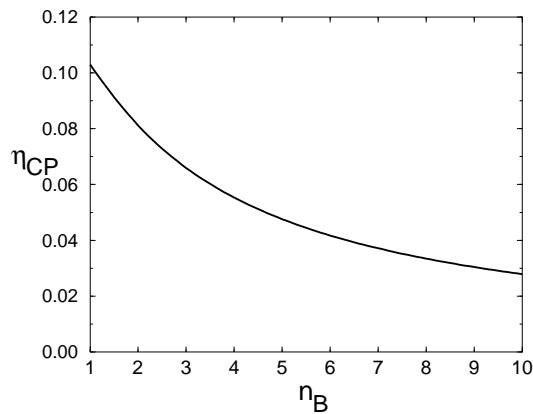,width=3.0in}
\end{center}
\end{figure}

In Fig. \ref{figrho} the volume fraction of the colloid
at the critical point is very low. The density range of the
colloidal liquid is very large, from the volume fraction at
the critical point, $0.048$, to $0.49$.
In Fig. \ref{figcps} we have plotted the
volume fraction of the colloid at the critical point, $\eta_{CP}$,
as a function of polymer size, $n_B$. For large $n_B$, it
decreases as $n_B^{-1}$ ($R_E^{-2}$). Thus it tends to
zero as $R_E\rightarrow\infty$ unlike the case for a pairwise
additive attraction where as its range ${\cal R}\rightarrow\infty$,
the volume fraction at the critical point tends to $0.13$.
The mixture phase
separates into colloid-rich (polymer--poor) and colloid-poor
(polymer-rich) phases at very low colloid concentrations when the
polymer is larger than the colloid. The reason for the $n_B^{-1}$
scaling is clear from Eqs. (\ref{rhos}) and (\ref{muex}).
The phase with the higher colloid density, the colloidal
liquid, must have a sufficiently high density that at fixed $z^*$ the
polymer density is significantly below that in colloidal vapour.
Now,  from Eq. (\ref{rhos}) we see that this requires a $\mu_{EX}$
which is at least of order unity (recall that $kT=1$)
in the liquid phase. At low colloid
density $\mu_{EX}=(A+Bn_B)\eta+O(\eta^2)$, where
$a$ and $b$ are constants. Thus, the colloidal
volume fraction at which $\mu_{EX}=1$
varies as $1/(A+Bn_B)$, which gives rise to a critical density
with the same scaling.
A critical volume fraction scaling as
$R_E^{-2}$ is consistent with work on a polymer molecule in the
presence of a density of fixed obstacles \cite{odijk00} which
finds that the reduction in entropy of the polymer molecule is of order
one when the number density of obstacles
of diameter $\sigma$ is of order $1/(\sigma R_E^2)$.

A further point to note is that as the activity of the polymer is
increased the density of the colloid-rich phase increases
rather slowly \cite{note}.
In Fig. \ref{figrho}, even when the polymer
activity is twice that at the critical point, the colloid
volume fraction in the colloid-rich phase is only around $0.15$.
Ultimately, we expect that if the polymer activity
is high enough there will be a triple point, where the dense fluid
is sufficiently dense that it coexists not only with a  dilute fluid
phase but with a crystalline phase. However,
this will be for much larger
polymer activities than shown in Fig. \ref{figrho}. The polymer density
in the colloid-poor, polymer-rich phase will be many times the
overlap concentration, $\rho^*=1$.
Although simultaneous coexistence of dilute and dense fluid phases
and a crystalline phase have been observed
in experiment for colloid + polymer
mixtures \cite{poon95}, this has been for values of $R_E$ no larger than
the colloid diameter. Observing simultaneous coexistence of these
three phases in experiment may be difficult if $R_E$ is
significantly greater than the colloid diameter.

In the limit of short polymers, $R_E$ a few tenths of $\sigma$
or less, the effect of polymer is to induce an attraction
which is short ranged, ${\cal R}\approx \sigma+R_E$, and to a good
approximation pairwise additive \cite{dijkstra99}. If the
phase diagrams for colloid + short polymer \cite{lekkerkerker92,meijer94},
and for particles with a short-ranged pairwise-additive attraction
\cite{hagen94}, are compared they are seen to be qualitatively the same.
In particular, in both cases as $R_E$ or ${\cal R}$ shrinks,
fluid-fluid coexistence disappears from the equilibrium
phase diagram. Thus, assuming that free polymer is equivalent in
effect to a pair attraction between colloidal particles
is a reasonable assumption for small
values of $R_E/\sigma$ but not for large values.

We have proposed an extended Asakura-Oosawa
model to model the interaction between colloidal particles and ideal
nonadsorbing polymer coils with end-to-end separations $R_E$ larger than the
diameter of the colloidal particle $\sigma$. For a globular protein with
diameter of a few nms, this would mean a polymer molecule with
an $R_E$ of 5nm or more.
As with smaller polymer coils
\cite{lekkerkerker92}, the polymer induces a vapour-liquid-like
separation into two fluid phases. One rich in colloid (a colloidal
`liquid') but poor in polymer and one poor in colloid (a colloidal
`gas') but rich in polymer. We showed that the critical point of this
transition moves to lower and lower colloid densities as the
polymer coils become larger. This is qualitatively different
from what would be found if the effect of polymer was to induce
a pairwise additive attraction between the colloidal particles.
Let us compare the variation of the colloid density
at a critical point induced by a pairwise additive attraction of varying
range ${\cal R}$, with that induced by
a polymer of varying size, $R_E$.
We find that for small ${\cal R}$ or $R_E$ the variation
of the critical density is similar in both cases. The critical density
increases \cite{lekkerkerker92,hagen94,sear00}
as ${\cal R}$ or $R_E$ shrinks; if not preempted by crystallisation
it tends to the random-close-packed density of hard spheres
as ${\cal R}$ or $R_E\rightarrow0$ \cite{sear00}.
However, in the other limit, that of large ${\cal R}$ or $R_E$,
the variation in the critical density is very different in the two cases.
With a pairwise additive potential it shrinks to $0.13$ \cite{vdw}
as ${\cal R}\rightarrow\infty$ and then goes
no lower, whereas with polymer coils, the present theory predicts
that the critical density tends to zero as $R_E\rightarrow\infty$.

Work supported by EPSRC (GR/N36981).

%\newpage

\end{multicols}

\newpage

%\clearpage

%\clearpage


\begin{thebibliography}{99}
\bibitem{wnote} Within TPT1 the Helmholtz free energy change on assembling
a chain of $n_B$ spheres is the number of contacts between
successive spheres, $n_B-1$, times the free energy change
on bringing a pair of spheres into contact from infinity, which
is, by definition, minus the logarithm of the
pair distribution function at contact. This is the origin of the
first term in the square brackets in Eq. (\ref{muex}); the
second term is simply $\eta$ times the density derivative of the
first term. This is required because we need the excess chemical
potential which is related to the excess Helmholtz free energy
per particle by $\mu_{EX}=a_{EX}+\eta(\partial a_{EX}/\partial\eta)$.
See Ref. \cite{sear94} for a discussion.
\bibitem{note} This is true even for the Asakura-Oosawa model with
$\sigma_P$ not much smaller than $\sigma$, as shown via computer
simulation by Meijer and Frenkel, Ref. \cite{meijer94}.
\end{thebibliography}
\end{document}